\documentclass[twocolumn,aps,prb,groupedaddress,showpacs]{revtex4}
\usepackage{amssymb}
\usepackage{amsmath}
\usepackage{bm}
\usepackage{graphicx}
\usepackage{color}

\begin{document}

\title{Long range triplet Josephson effect through a ferromagnetic trilayer}
\author{M. Houzet$^1$ and A. I. Buzdin$^2$}
\affiliation{$^1$ DRFMC/SPSMS, CEA Grenoble, 17, rue des Martyrs, 38054 Grenoble Cedex 9,
France\\
$^2$ Institut Universitaire de France and Universit\'{e} Bordeaux I, CPMOH,
UMR 5798, 33405 Talence, France}
\date{\today}

\begin{abstract}
We study the Josephson current through a ferromagnetic trilayer, 
both in the diffusive and clean limits. For colinear (parallel or
antiparallel) magnetizations in the layers, the Josephson current is small
due to short range proximity effect in superconductor/ferromagnet
structures. For non colinear magnetizations, we determine the conditions for
the Josephson current to be dominated by another contribution originating
from long range triplet proximity effect.
\end{abstract}

\pacs{
74.45.+c, 74.78.Fk,  85.25.Cp. }
\maketitle

The coexistence of superconductivity and ferromagnetism is very rare in the
bulk systems. However, it can be easily achieved in the artificially
fabricated superconductor/ferromagnet (S/F) heterostructures. The S/F
proximity effect is characterized by the damped
oscillatory behavior of the Cooper pair wave function in the ferromagnet. This
phenomenon leads to the non-monotonous dependence of the critical
temperature of S/F multilayers on the F layer thickness and the
realization of the Josephson $\pi$-junctions (for a review see 
Refs.~\onlinecite{Buzdin-rew,GolubovRMP}). In the diffusive limit, the proximity
effect in F metal is rather short-ranged due to the large value of the
ferromagnetic exchange field. This is related with the
incompatibility between singlet superconductivity and ferromagnetism.

Interestingly, the non-uniform magnetization can induce the triplet
superconducting correlations which are long-ranged (on the same scale
as for superconductor/normal (N) metal proximity effect) 
\cite{BergEfVolk-rew}. It exists several experimental indications on this triplet
proximity effect \cite{Sosnin,Xiao}. However, the transition from usual to
long range triplet proximity effect was never observed in the same system.

\begin{figure}
\includegraphics[width=55mm]{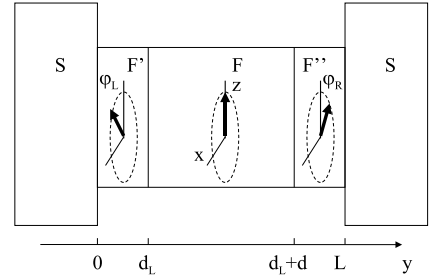}
\caption{Geometry of S/F'/F/F''/S junction. The arrows indicate non-colinear
orientations of magnetizations in each layer with thickness 
$d_L$, $d$, $d_R$, respectively ($L=d_L+d+d_R$). }
\label{F:fig1}
\end{figure}

In the present work, we investigate the conditions for the observation of
the Josephson current due to a long range triplet component under controllable
conditions. The non-colinear magnetization may serve as a source of the
long range triplet component. However, it is not possible to have the
Josephson current due to the interference of the triplet and singlet
components. Two sources of the triplet components are needed to observe the
long range triplet Josephson effect between them. Then, the simplest
experimental realization of such a situation may be the S/F'/F/F''/S system
with the magnetic moments of the F', F'' layers non-colinear with the F
interlayer (see Fig.~\ref{F:fig1}). The optimal condition for the triplet
Josephson current observation is when the thicknesses $d_L$ and $d_R$ of the
layers F', F'' are of the order of the coherence length $\xi_{f}$ in the
ferromagnet. Indeed, for large $d_{L},d_R$, the triplet component is exponentially
small due to short range proximity effect in the layers F' and F'', while for 
very thin $d_{L},d_R$, it is also small. Then, we predict that the 
magnitude of the Josephson current in the structure with F layer 
thickness much larger than $\xi_f$ will be comparable to that of an S/N/S junction 
with the same length. 

The similar phenomenon could be observed in lateral Josephson junctions
made of a nanostructured ferromagnetic film allowing control
on its magnetic domain structure. Then, the described effect
would give a much larger critical current than the one predicted in S/F/S junctions 
with in-plane magnetic domain walls.\cite{fominov}

Besides, the triplet Josephson effect provides the possibility of the
0 and $\pi $-junction realization due to the different orientations of
magnetic moments in the F' and F'' layers. Such effect was
revealed in S$_{\text{F}}$/I/S$_{\text{F}}$ junctions where S$_{\text{F}}$
are magnetic superconductors with helical magnetic order separated by a thin
insulating (I) layer\cite{Kulic}. It was also obtained in diffusive F/S multilayers
with non colinear magnetizations in successive F layers \cite{bergeret}. In
this case, the triplet Josephson effect is mediated by the inverse proximity
effect in the thin S layers. It would compete with the reduction of
critical temperature and gap amplitude, but these were not taken into
account. In Ref.~\onlinecite{nazarov}, an idealized circuit-theory model for
the triplet proximity effect in an S/F/I/F/I/F/S junction was proposed. The
spatial range of singlet and triplet proximity effect was not considered.
Our work is somewhat complimentary to these ones. The question about the
concrete realization and optimization of the triplet Josephson effect was
outside the scope of these approaches while it is of primary importance in
the present study. 

We also provide an analysis of the triplet Josephson current in 
the ballistic (clean) limit. In this case, the singlet component reveals the
non-exponential oscillatory decay but nevertheless the decay of the triplet
component is even weaker and it is again possible to observe the crossover
between singlet and triplet Josephson effects. 

The needed conditions for the triplet proximity effect observation in the Josephson
current are rather stringent. The considered system, if
realized experimentally, could provide an excellent opportunity to study the
crossover between the triplet and singlet Josephson effects with the
rotation of the magnetic moment of any of the F layers.

Let us now calculate the supercurrent through a Josephson junction made of a
ferromagnetic trilayer attached to superconducting leads, according to the 
geometry depicted in Fig.~\ref{F:fig1}. We assume that the layers are in good
electric contact and that the magnetizations in the layers have the same 
amplitude. The exchange field $\bm{h}$ acting on the spin of the conduction 
electrons is parallel to the magnetizations, with following spatial
dependence:
\begin{equation}\label{eq:mag-struct}
\bm{h}(y)=\left\{ 
\begin{array}{lr}
h(\sin\phi_L\hat{x}+\cos\phi_L\hat{z}), & 0<y<d_L, \\ 
h\hat{z}, & d_L<y<d_L+d, \\ 
h(\sin\phi_R\hat{x}+\cos\phi_R\hat{z}) , & d_L+d<y<L,%
\end{array}
\right.
\end{equation}
where $d_L$, $d$, and $d_R$ are the thicknesses of each layer, and $L=d_L+d+d_R$
is the total length of the junction. Here we adopt the sames axes for space
and spin quantization.

We first consider the diffusive limit, when the mean free path is shorter
than the widths of the layers and coherence lengths. For simplicity, we also 
assume that the temperature is close to the critical temperature of the leads.
Then, within the quasiclassical theory of superconductivity \cite{Usadel},
the current flowing through the junction is
\begin{equation}
I=\frac{G L}{e}\pi T\sum_{\omega >0}\mathrm{Im}\mathrm{Tr}[\hat{F}^{\ast }(y)%
\hat{\sigma}_{y}\hat{F}^{\prime }(y)\hat{\sigma}_{y}],  \label{eq:courant}
\end{equation}
where the anomalous Green's function $\hat{F}=F_{0}+\bm{F}.\hat{\bm{\sigma}}$
is a matrix in spin space and it solves the linearized Usadel equation in the
ferromagnet: 
\begin{equation}
-D\hat{F}^{\prime \prime }(y)+2\omega \hat{F}(y)+i\bm{h}(y).\{%
\bm{\hat{\sigma}},\hat{F}(y)\}=0  \label{eq:usadel}
\end{equation}
(in units with $\hbar=k_{B}=1$). Here, $G $ is the conductance of the
junction in its normal state, $D$ is the diffusion constant of the 
ferromagnet, $\omega =(2n+1)\pi T$ are the Matsubara frequencies at temperature $T$, $%
\hat\sigma _{i(i=x,y,z)}$ are the Pauli matrices, and the primes denote
derivative along $y$-direction.
Depairing currents generated by the orbital effect have been neglected
in eq.~(\ref{eq:usadel}),
as usually done for ferromagnetic layers with in-plane 
magnetization\cite{Buzdin-rew}.

The Usadel equation (\ref{eq:usadel}) is solved in the central F layer in
terms of its values at the interfaces with F' and F'' layers: 
\begin{eqnarray}  \label{eq:sol-central}
&&F_0(y)\pm F_z(y)=\left[F_0(d_L)\pm F_z(d_L)\right]\frac{\mathrm{sh}
q_\pm(d_L+d-y)}{\mathrm{sh} q_\pm d}  \notag \\
&& \qquad \qquad +\left[F_0(d_L+d)\pm F_z(d_L+d))\right]\frac{\mathrm{sh}
q_\pm(y-d_L)}{\mathrm{sh} q_\pm d},  \notag \\
&&F_x(y)=F_x(d_L)\frac{\mathrm{sh} q_0(d_L+d-y)}{\mathrm{sh} q_0d}  \notag \\
&& \qquad \qquad +F_x(d_L+d)\frac{\mathrm{sh} q_0(y-d_L)}{\mathrm{sh} q_0d},
\end{eqnarray}
and $F_y=0$, as $\bm{h}$ has no component along $\hat{y}$-direction. Here, $q_0=%
\sqrt{2\omega/D}$ and $q_\pm=\sqrt{2(\omega\pm ih)/D}$. As the amplitude of
exchange field is much larger than critical temperature $T_c$, we may
simplify $q_\pm\simeq(1\pm i)/\xi_f $, where $\xi_f =\sqrt{D/h}$ is the
ferromagnet coherence length and is much shorter than superconducting
coherence length $\xi_0 =\sqrt{D/2\pi T_c}$. The solutions of eq.~(\ref%
{eq:usadel}) in the other layers, as well as their derivative, should match
continuously eq.~(\ref{eq:sol-central}) at each interface. In absence of
interface barriers with the S leads, they should also take the values $\hat{F%
}(y=0,L)=\hat{F}^{L,R}$, where $\hat{F}^{L,R}=(\Delta/\omega) e^{\mp i\chi/2}
$ are bulk solutions in the leads. Here, $\Delta$ is the modulus of the
superconducting gap and $\chi$ is the phase difference maintained between
the leads. Close to $T_c$, the gap vanishes as $\Delta(T)=[(8\pi^2/7%
\zeta(3))k^2T_c(T_c-T)]^{1/2}$. Here, we neglect selfconsistency for the gap
equation in the leads, as usually done assuming that the width of S
electrodes is much larger than that of F layers, or that the Fermi velocity
in F layers is smaller \cite{GolubovRMP}.

To proceed further with tractable formulas, we assume that F' and F'' layers are
thin: $d_L,d_R\ll\xi_f $. Then, the solution
in F' layer varies only slightly with $y$ and can be put in approximate
form: 
\begin{eqnarray}  \label{eq:formF}
\hat{F}(y)&\simeq&\hat{F}(d_L)+(y-d_L)\hat{F}^{\prime}(d_L)  \notag \\
&&-\frac{(y-d_L)^2}{d_L^2}[\hat{F}(d_L)-d_L\hat{F}^{\prime}(d_L)-\hat{F}^L],
\end{eqnarray}
which satisfies the boundary conditions at $y=0$ and $y=d_L$. In addition,
it should also solve the Usadel equation. Inserting eq.~(\ref{eq:formF}) into (%
\ref{eq:usadel}), we get: 
\begin{equation}  \label{eq:interm-bound}
\frac{D}{d_L^2}[\hat{F}(d_L)-d_L\hat{F}^{\prime}(d_L)-\hat{F}^L]+ \frac{i}{2}
\bm{h}.\{\bm{\hat\sigma},\hat{F}^L\} \simeq 0,
\end{equation}
where a term $\omega \hat{F}^L$ was neglected (as $h\gg T$). Equation (\ref%
{eq:interm-bound}) yields the results: 
\begin{subequations}
\begin{eqnarray}
F_0(d_L)&=&F_0^L, \\
F_x(d_L)&=&-i(d_L^2 h/D)\sin\phi_L F_0^L, \\
F_z(d_L)&=&-i(d_L^2 h/D)\cos\phi_L F_0^L,
\end{eqnarray}
provided that $d_L|\hat{F}^{\prime}(d_L)|\ll |\hat{F}(d_L)|$, as can be
checked consistently from eq.~(\ref{eq:sol-central}) when $d_L\ll\xi_f$.

Similar results can be obtained for $\hat{F}(y=d_L+d)$. We can now evaluate
eq.~(\ref{eq:courant}), say at $y=d_L$, and we find $I=I_c \sin\chi$, where
the critical current is: 
\end{subequations}
\begin{equation}  \label{eq:Icdiff}
I_c \!=\!\frac{2 \pi TG }{ e} \!\sum_{\omega>0} \!\frac{\Delta^2}{\omega^2}
\!\left\{\! \mathrm{Re}\frac{q_+d}{\mathrm{sh} q_+d}\! -\!\frac{q_0d}{%
\mathrm{sh} q_0d}\frac{d_L^2d_R^2}{\xi_f^{4}}\sin\phi_L\sin\phi_R \!\right\}.
\end{equation}

The first term in eq.~(\ref{eq:Icdiff}) comes from short range singlet ($%
F_{0}$) and triplet ($F_{z}$) components of anomalous function $\hat{F}$. It
equals the critical current of an S/F/S junction with length $d$.%
\cite{BuzKup1991} Its sign oscillates with varying ratio $d/\xi_{f} $. In
particular, when $d\gg \xi _{f}$, 
\begin{equation}  \label{eq:ic-sfs}
I_{cf} =\frac{\pi G }{2\sqrt{2}e}\frac{\Delta(T)^{2}}{T_{c}}\frac{d}{\xi
_{f} }\sin \left( \frac{\pi }{4}+\frac{d}{\xi _{f} }\right) e^{-d/\xi _{f} }.
\end{equation}%
Thus, its amplitude is also exponentially suppressed.

\begin{figure}
\includegraphics[width=62mm]{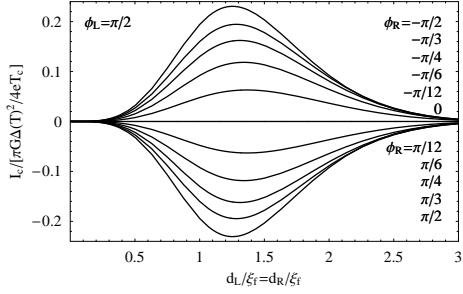}
\caption{ Critical current induced by long range triplet proximity effect in
S/F'/F/F''/S junction, in units of $(\protect\pi G \Delta(T)^2/4eT_c)$, for
varying length of F' and F'' layers, at $d_L=d_R\sim \protect\xi_f\ll d\ll%
\protect\xi_0$, and for different orientations of the magnetization in the
layers. }
\label{F:fig2}
\end{figure}

The second term in eq.~(\ref{eq:Icdiff}) comes from long range triplet
component ($F_{x}$) and yields:
\begin{equation}\label{eq:ict-dirty}
I_{ct}=-I_{cn}(d_L^2d_R^2/\xi_f^4)\sin\phi_L\sin\phi_R,
\end{equation}
where $I_{cn}$ is the critical current in S/N/S junction:\cite{GolubovRMP} 
\begin{equation}
I_{cn}=\frac{G}{e}2\pi T\sum_{\omega >0}\frac{q_{0}d}{\mathrm{sh}q_{0}d}%
\frac{\Delta ^{2}}{\omega ^{2}}.
\end{equation}
In particular, in junctions with length $d\ll \xi _{0}$: $I_{cn}=(\pi
G\Delta ^{2}/4eT_{c})$. The small prefactor $(d_{L}^{2}d_{R}^{2}/\xi
_{f}^{4})$ in eq.~(\ref{eq:ict-dirty}) comes from the simplifying assumption 
$d_{L},d_{R}\ll \xi _{f}$ that we used in the calculation. As explained 
in Introduction, $I_{ct}$ would be reduced by the exponential factor 
$e^{-(d_{L}+d_{R})/\xi _{f}}$at $d_{L},d_{R}\gg \xi _{f}$. Thus, at
optimal size $d_{L},d_{R}\sim \xi _{f}$, the second term 
$I_{ct}\sim -I_{cn}\sin \phi _{L}\sin \phi _{R}$ is much
larger than the first one, $I_{cf}$, provided that the magnetic layers have
non colinear orientations. For arbitrary lengths $d_{L},d_{R}\sim \xi _{f}$,
the critical current originating from long range triplet correlation only,
at $\xi _{f}\ll d\ll \xi _{0}$, was also obtained from eqs.~(\ref{eq:courant}%
) and (\ref{eq:usadel}), see Fig~\ref{F:fig2}. We see that the
triplet contribution to the critical current may be  observed on the
experiment only in the rather small interval of the F', F''
layers thickness: $d_L,d_R\sim (0.5-2.5)\xi _{f}$.

The dependence of the critical current with the orientations of the
magnetizations in successive F layers similar to eq.~(\ref{eq:Icdiff}) was
obtained in Refs.~\onlinecite{bergeret,nazarov}. We note that the sign of
the long range component of critical current can be tuned with these
orientations. This component is absent in the case of only two layers with
opposite \cite{hekking}, or even non-colinear magnetizations\cite{crouzy}.

The Usadel equations would easily allow generalizing the result (\ref%
{eq:Icdiff}) obtained here. (i) Qualitatively, the above result should not rely on the 
assumption that the temperature is close to $T_c$ and it would be preserved 
even at smaller temperature. (ii) Barrier interfaces
between the layers and the leads would
decrease both short range and long range contributions to critical current.%
\cite{nazarov} (iii) Equation (\ref{eq:mag-struct}) may also describe
the case of a ferromagnet with magnetic domains and thin domain walls 
(few atomic lengths). If the domain walls are large, the
long range triplet contribution will be decreased by the factor 
$\xi_f /\delta_w\ll 1$, where $\delta_w$ is the domain wall width, in
analogy with the theory of enhanced critical temperature in S/F bilayers
due to domain-wall superconductivity \cite{DWS}. 

Note an interesting possibility to separate the triplet and singlet
Josephson effects even for relatively thin central F layer $d\sim \xi _{f}$.
Indeed if its thickness is around the first critical value $(3\pi/4)\xi _{f}$%
, see eq.~(\ref{eq:ic-sfs}), the temperature variation may serve as a fine
tuning and provoke the $0/\pi$ transition \cite{RyazanovPRL01,Oboznov}. For
the S/F'/F/F''/S system, the singlet component would vanish at such
temperature and only the triplet critical current would be observed.

We consider now the clean limit. Then, the supercurrent flowing through the
junction is now given by: 
\begin{equation}
I=-\frac{2\pi TG }{e}\sum_{\omega >0}\int \frac{d\Omega _{\bm{n}}}{4\pi }%
n_{x}\mathrm{Im}\mathrm{Tr}\left[ \hat{f}_{-\bm{n}}^{\ast }(y)\hat\sigma _{y}%
\hat{f}_{\bm{n}}(y)\hat\sigma _{y}\right] ,  \label{eq:current-clean}
\end{equation}%
where $\hat{f}_{\bm{n}}(y)$ solves the Eilenberger equation in F layer: 
\begin{equation}
\bm{v}.\bm{\nabla}\hat{f}_{\bm{n}}(y)+2\omega \hat{f}_{\bm{n}}(y)+i\bm{h}%
.\left\{ \bm{\hat\sigma},\hat{f}_{\bm{n}}(y)\right\} =0.  \label{eilenberger}
\end{equation}%
Here, $\bm{v}=v\bm{n}$ is a Fermi velocity, $\bm{n}$ is a unit vector, $G$
is the Sharvin conductance of the ballistic junction in its normal state. In
addition, the solution of eq.~(\ref{eilenberger}) should be continuous, and
match with the bulk solution in S lead where the electrons come from. That is, $%
\hat{f}_{\bm{n}}(y=0)=\hat{F}^{L}$ if $n_{y}>0$, $\hat{f}_{\bm{n}}(y=L)=\hat{%
F}^{R}$ if $n_{y}<0$. Again, we neglect selfconsistency for the gap
equation in the leads.

Solving eq.~(\ref{eilenberger}) at $0<y<d_{L}$ and $n_{y}>0$, we find for $%
\hat{f}_{\bm{n}}(y)\equiv f_{0}+\bm{f}.\hat{\bm{\sigma}}$ that: 
\begin{eqnarray}
f_{0}\pm (\sin \phi _{R}f_{x}+\cos \phi _{R}f_{z}) \!&=&\! (\Delta/\omega
)e^{-i\chi /2}e^{-2(\omega \pm ih)y/v_{y}},  \notag  \label{eq:sol-left} \\
\sin \phi _{R}f_{z}-\cos \phi _{R}f_{x} \!&=&\!0.
\end{eqnarray}%
Then, using continuity of $\hat{f}$ at $y=d_{L}$ and solving eq.~(\ref%
{eilenberger}) at $d_{L}<y<d_{L}+d$, we find: 
\begin{eqnarray}
f_{0}\pm f_{z} &=&\alpha e^{-2(\omega \pm ih)(y-d_{L})/v_{y}}(c_{d_{L}}\mp
is_{d_{L}}\cos \phi _{L}),  \notag \\
f_{x} &=&-i\alpha \sin \phi _{L}s_{d_{L}}e^{-2\omega (y-d_{L})/v_{y}}
\end{eqnarray}%
where $\alpha =(\Delta /\omega )e^{-i\chi /2}e^{-2\omega d_{L}/v_{y}}$, and
we use short notations $s_{d_{L}}=\sin (2hd_L/v_{y})$, $c_{d_{L}}=\cos
(2hd_L/v_{y})$. Similar solution can be found for $\hat{f}$ at $n_{y}<0$.
The supercurrent (\ref{eq:current-clean}) is then conveniently evaluated at $%
y=d_{L}+d/2$ and we find $I=I_{c} \sin \chi $, where: 
\begin{eqnarray}
I_{c} &=&\frac{4\pi TG }{e}\sum_{\omega >0}\int_{0}^{1}dn_{y}n_{y}\frac{%
\Delta ^{2}}{\omega ^{2}}e^{-\frac{2\omega L}{v_{y}}}\left[
c_{d}c_{d_{L}}c_{d_{R}}\right.  \notag \\
&-&c_{d}s_{d_{L}}s_{d_{R}}\cos \phi _{L}\cos \phi
_{R}-s_{d}c_{d_{L}}s_{d_{R}}\cos \phi _{R}  \notag \\
&-&\left. s_{d}s_{d_{L}}c_{d_{R}}\cos \phi _{L}-s_{d_{L}}s_{d_{R}}\sin \phi
_{L}\sin \phi _{R}\right] .
\end{eqnarray}%
To proceed further, we assume that $d_{L},d_{R}\ll \xi _{f} \ll d$, where the 
ferromagnet coherence length $\xi _{f} =v/h$ in clean limit is
much shorter than superconducting coherence length $\xi _{0} =v/2\pi T_{c}$.
Then, 
\begin{eqnarray}
I_{c} &\simeq &\frac{4\pi TG }{e}\sum_{\omega >0}\int_{0}^{1}dn_{y}n_{y}%
\frac{\Delta ^{2}}{\omega ^{2}}e^{-\frac{2\omega d}{v_{y}}}\left[ \cos
\left( \frac{2hd}{v_{y}}\right) \right.  \notag \\
&&-\left. \sin \left( \frac{2hd_{L}}{v_{y}}\right) \sin \left( \frac{2hd_{R}%
}{v_{y}}\right) \sin \phi _{L}\sin \phi _{R}\right] .
\end{eqnarray}
Here, the first term comes from short range proximity effect. It coincides
with the critical current of clean S/F/S junction with length $d$. In
particular, at $\xi _{f} \ll d\ll \xi _{0} $, it yields \cite{BuzBul82}: 
\begin{equation}
I_{cf} =-\frac{\pi \Delta ^{2}G }{2eT_{c}}\frac{\xi _{f} }{2d}\sin 
\left( \frac{2d}{\xi _{f} }\right) .
\end{equation}
The second term comes from long range triplet proximity effect and yields
(for $d_{L}\sim d_{R}\ll\xi_f\ll d\ll \xi_0$): 
\begin{equation}
I_{ct} =-\frac{\pi \Delta ^{2}G }{2eT_{c}}\left[ \frac{4d_{L}d_{R}}{\xi
_{f}^2}\ln \frac{\xi _{f} }{2(d_{L}+d_{R})}\right] \sin \phi _{L}\sin \phi
_{R}.  \label{eq:Ictriplet}
\end{equation}%
It is small under assumption $d_{L},d_{R}\ll \xi _{f} $. On the other hand,
at $d_{L},d_{R}\gg \xi _{f} $, the critical current (\ref{eq:Ictriplet})
would be suppressed by the factor $\xi_f^2/d_Ld_R\ll 1$, due to short range
proximity effect in F' and F'' layers. Again, we expect a maximum of
critical current at $d_{L}\sim d_{R}\sim \xi _{f} $, with amplitude $%
I_{ct}\propto -I_{cn}\sin\phi _{L}\sin \phi _{R}$, where $I_{cn}=(\pi \Delta
^{2}G /4eT_{c})$ is the critical current of a clean S/N/S junction with $%
d\ll \xi _{0}$. The dependence of the critical current on the orientations of 
the magnetizations in F layers is similar to the diffusive case.

The Josephson current through a half-metal (HM) with one spin band only is
expected to vanish\cite{Xiao,Eschrig}. However, spin-flip processes taking
place at S/F interfaces were suggested to promote triplet correlation and
induce a finite supercurrent through the device\cite{Eschrig,Tanaka,Eschrig2}%
. The quasiclassical theory presented in this work assumes that
ferromagnetic exchange field is much smaller than the Fermi energy.
Therefore, it is not well suited to address quantitatively the case of HMs,
when they are comparable. Qualitatively, the non colinear layers F' and F''
with thicknesses of the atomic scale would play the role of spin flip
scatterers with inverse scattering time $\tau_\text{sf}^{-1}$ proportional
to spin band splitting $h$. Then, the order of magnitude for the triplet
induced supercurrent can be obtained from eq.~(\ref{eq:Ictriplet}) by noting
that the reduction factor $d_Ld_R/\xi_f^2$ (up to the log term) is
proportional to $1/(\tau_\text{sf}E_F)^2$, where $E_F$ is Fermi energy. It
is thus proportional to the probability for an electron from the minority
spin band to be transfered through HM by spin-flip processes at the interfaces
with the leads.

In conclusion, we determined the Josephson current through a ferromagnetic
trilayer. For colinear (parallel or
antiparallel) magnetizations in the layers, the Josephson current is small
due to short range proximity effect in superconductor/ferromagnet
structures. For non colinear magnetizations, we determined the conditions for
the Josephson current to be dominated by another contribution originating
from long range triplet proximity effect. In practice the triplet
Josephson current may be observed in systems with the lateral layers 
thickness of the order of $\xi _{f}$ only.

The studied structures offer an interesting possibility to study the
interplay between Josephson current and dynamic precessing of the magnetic
moment. Indeed we may expect the strong coupling between ferromagnetic
resonance (or/and spin waves) and Josephson current - in particular the
additional harmonics generation in ac Josephson effect.

We acknowledge Norman Birge and Louis Jansen for a critical reading of the
manuscript and useful comments.

\end{document}